# [Preprint] High-Dimensional Data Classification in Concentric Coordinates

Alice Williams
Department of Computer Science
Central Washington University
USA
0009-0001-6154-2407

Boris Kovalerchuk
Department of Computer Science
Central Washington University
USA
0000-0002-0995-9539

*Abstract*—The visualization of multi-dimensional data with interpretable methods remains limited by capabilities for both high-dimensional lossless visualizations that do not suffer from occlusion and that are computationally capable by parameterized visualization. This paper proposes a low to high dimensional data supporting framework using lossless Concentric Coordinates that are a more compact generalization of Parallel Coordinates along with former Circular Coordinates. These are forms of the General Line Coordinate visualizations that can directly support machine learning algorithm visualization and facilitate human interaction.

*Keywords*—*Multi-Dimensional Data Visualization, Lossless Visualization, Concentric Coordinates, Circular Coordinates, Visual Knowledge Discovery, AI/ML Data Classification.*

## I. INTRODUCTION

### A. Motivation

In many domains, accurate and interpretable classification models can be accurately visualized. However, in many other domains, this remains a long-standing and critical roadblock to deploy artificial intelligence and machine learning (AI/ML) models. This is critical and challenging for high-risk tasks like healthcare diagnostics. Visualization of multidimensional (n-D) data classification is critical for three major reasons:

(1) to speed up analysis of prediction accuracy,
(2) to interpret/explain classifier predictions, and
(3) to improve/modify the prediction model.

### B. Overview of Existing Methods

AI/ML tasks for high multi-dimensional (n-D) data are commonly approached with black-box deep-learning (DL) methods that inherently lack in interpretability and decision explanation. Further relying on explainability after model design as popularly done with either LIME or SHAP [7]. Moreover, visualization methods used commonly preprocess data with dimensional reduction (DR) methods like Principal Component Analysis (PCA), t-Stochastic Neighbor Embedding (t-SNE), or other similar approximations. However, such methods are lossy and not reversible. Therefore, these methods commonly introduce visually verify inaccuracies in n-D. Alternatively, lossless visualizations allow for the use of Visual Knowledge Discovery (VKD) to visually discover algorithmic adjustments that improve ML prediction models [5].

### C. High-Dimensional Challenges

The difficulty to classify n-D data is complicated by the dimensionality present in many real-world datasets. Fig. 1 shows how the popular DR method of t-SNE can alter and misrepresent class boundaries additionally introducing artificial outlier cases. Thus, while data label classes do form visual clusters, even in this lossy visualization, there are added outliers frequently placed inside each class cluster. This shows that clusters are falsified due to not always properly respecting n-D case order. Examples of such cases are circled in Fig. 1 in each class cluster of the MNIST 10 class 784-D data. This introduces severe classification limitations that are problematic for classical classification algorithms like k-Nearest Neighbors (k-NN). So, while we can visualize n-D cases, the highly dimensional cases are frequently placed improperly in visualizations as known by the Johnson-Lindenstrauss Lemma [15] [8] that bounds the distortion in mapping n-D points to 2-D.

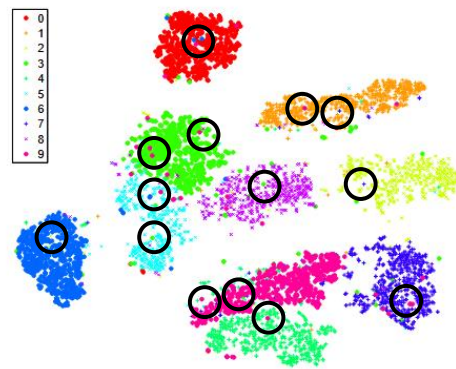

Fig. 1. Scatterplot visualization of the 60,000 training cases from the MNIST dataset processed with the popular t-SNE dimensional reduction method [15].

## II. VISUALIZATION IN CONCENTRIC COORDINATES

### A. Background

Concentric Coordinates (CoC) were originally introduced in [19] to visualize computer network scans and traffic data. This paper generalizes the method as a new General Line Coordinate (GLC) lossless visualization that is directly related to the previous Static and Dynamic Circular Coordinates (SCC, DCC) [18]. The CoC generalizes circle axis-based visualization methods like the TimberTrek [15] approach for visualization of decision trees (DT) on a circle. This paper extends the CoC visualization to visualization of n-D ML/AI data with higher dimensionality and for predictive ML model discovery.

CoC is descendant from both Shifted Paired Coordinates Decision Trees (SPC-DT) [9] and Elliptical Coordinates [12]. Moreover, the CoC generalizes Parallel Coordinates [8] by visualizing n-D data with either a single, or multiple circle axes per attribute visualized. So, the visualization choice can support different data dimensionalities. Additionally, n-D cases can be visually highlighted across the attribute circles like in Bended Coordinates and SPC-DT [9]. Additionally, attribute reordering visually affects the data geometry, as suggested in foundational works for pattern recognition using the Parallel Coordinates [6].

Below, we contrast **Circular Coordinates** (**CC**) [18] and **Concentric Coordinates** (**CoC**), both are forms of the General Line Coordinates (GLC) [8] and are useful for visualization of n-D data. However, in CC **each circle** is used to visualize **a single class** of n-D data with **all attributes** of an n-D point placed on this circle. Fig. 2a demonstrates the CoC for all three Fisher Iris data classes. Here each n-D point is visualized as a sequence of end-point connected curves of 4 points on the circle for the class constructing 4-D point representations. For two of the data classes CC can use a single circle without introducing any occlusion by placing one class in the inner space of the circle and another class on the outer space of the circle. This is shown in Fig. 2a with red (Setosa) and green (Versicolor) classes on one circle and the blue (Virginica) class being on another circle.

Respectively, in CC the user can capture class discrimination patterns visually by exploring sectors of the circle(s) that contain cases of a single class, as is shown by two black lines from the common center of the circles in Fig. 2a. Thus, CC are efficient when such segments exist and can be visibly demarcated without any additional occlusion like is shown for the Iris data in Fig. 2a. However, in many classification tasks while such patterns do exist, they can only be made visible for some **subsets** of the data. So, revealing these patterns requires other visualization methods for the same data. Those other methods likely also will reveal visual patterns only for the subsets of data, which shows the need to use several visualization methods to reveal patterns fully [8] This is in line with several different forms of GLC that were proposed [10], this paper develops the Concentric Coordinates.

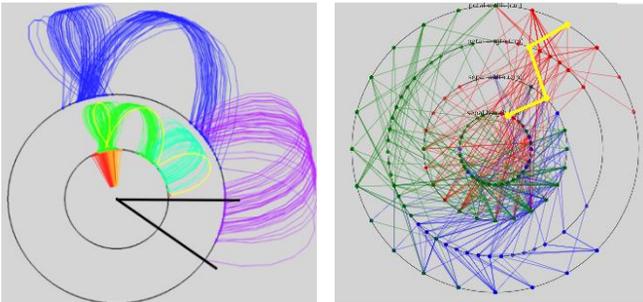

(a) Dynamic Circular Coordinates. Outer circle shows blue class, inner circle shows red class (outside) and green class (inside). One green class case is highlighted yellow.

(b) Concentric Coordinates so each circle is one attribute, each class is shown by class-colored polylines over all circles, one red class case is highlighted yellow.

Fig. 2. Three classes of 4-D Fisher Iris data visualized in Circular Coordinates (a) and Concentric Coordinates (b).

### B. Concept of Concentric Coordinates

**Concentric Coordinates** (**CoC**) use an individual circle axis **per attribute.** This requires using *n* circles to represent all dimensions of n-D datapoints. The **number of concentric circles** is a fundamental limitation of the CoC because we can only draw a limited number of circles on the screen at one time. Also, humans have limited perceptual capabilities to analyze a large number of concentric coordinates. In contrast, the CC uses only one circle per one or two classes of as shown in Fig. 1a.

The limitation of CC is the **number of attributes** that can be located on a single circle and can be observed for analysis by a human. This shows that circular and concentric coordinates complement each other in terms of visualization capabilities.

By default, in CoC the first coordinate $X_1$ is closest to the center, coordinate $X_2$ is the next one, and so on up to $X_n$. None of the coordinates cross each. The change of the attribute order can reveal certain patterns better. Also, by default, the start of each coordinate is in the north point as it is in circular and elliptic coordinates [12]. For n-D point $\mathbf{x} = (x_1, x_2, …, x_n)$ the values $x_i$ are mapped to points of respective coordinates $X_i$ with the *clockwise* distance from the north point equal to $x_i$, see Fig. 1b. These points of coordinates circles are connected by straight or curved lines. Circles can be rotated to reveal patterns better.

The CoC are a direct generalization of Parallel Coordinates with the change being substitution of parallel line axes to concentric circles. The CoC were introduced in [19] with an application in visual analytics of computer network data. This let authors conduct security visual analytics by observing data state transitions [3] in detecting *unusual patterns* of network scans, port scans, hidden scans, DDoS attacks, etc. Thus, this study was focused primarily on visually discovering **outliers**. This paper generalizes CoC as a form of the General Line Coordinates (GLC) for general ML c**lassification tasks** of data with low to high dimensionality. This paper explores novel variations on CoC using computational integration made visual through methods including density/opacity polyline weighting, visual tuning of circle rotation, and various generalized forms like planar rearrangements, and 3-D stacked visualizations.

The CoC version as originally proposed in [19] used parameterized curves to connect n-D points over the concentric circles. In comparison with the parallel coordinates' method, it decreased crossings of curves by over 15% due to the ability to strategically build curves in two opposing directions. This advantage is not available in parallel coordinates, however, while it is an advantage it was often insufficient to reveal the patterns hidden by visual occlusion. We propose additional methods to parameterize axes instead that further decrease the occlusion. There is also a similar circle axis based TimberTrek approach [16] to decision tree (DT) visualization that may benefit from further development of CoC.

Generally, in CoC each coordinate is a circle, ellipse, or **any convex closed contour** around a common center point **c** in **2-D** visualization. In section IV, we generalize the CoC to **multiple CoC** axes and extend visualization to **3-D**. Respectively, in 3-D visualization of the closed contours are spheres, ellipsoids, or any convex 3-D shapes topologically equivalent to the sphere. To represent a DT in the CoC we use the idea presented in the TimberTrek [16] visualization where polyline segments are marked with the class when it reaches the terminal nodes at the other circle/coordinate. TimberTrek focuses on visualizing pure DT flow, while we visualize losslessly the entire ML/AI n-D dataset reversibly. Moreover, in TimberTrek users can visualize subsets of the data, however, it is not a lossless visualization. Fig. 3 shows the difference in visualization of 4-D Iris data for Iris classes visualized separately in the same CoC clearly presenting the difference in their patterns.

### C. Plotting of Concentric Coordinates

Figs. 3-8 show that CoC allows a user to interactively tune the visualization of cases to be simpler or in a visual form that is more perceptually recognizable, such as closed polygons. This can simplify and reveal patterns visually since the CoC axes/circles can be **resized**, **rescaled**, **reordered**, and **rotated** to best reveal visual patterns.

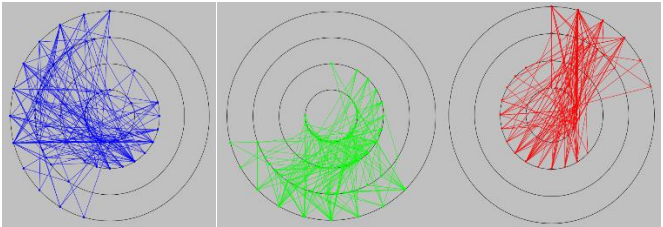

(a) Virginica class.   (b) Versicolor class.   (c) Setosa class.
Fig. 3. Three Fisher Iris 4-D data classes visualized in individual CoC plots.

Studies in perception have shown that a closed contour has an advantage for visual identification [6], however, it increases occlusion for the CoC. Therefore, the software allows a user to toggle visualization of the CoC with the closed contour. Figures shown in this section use closed contours and Fig. 4b shows an open contour. This shows that since the closed contour can bend and contradict the flow of attribute axes radiating outwards, it does not immediately help visual analysis like shown in Fig. 4a.

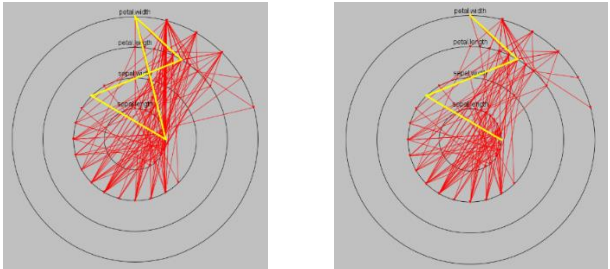

(a) Closed contour visually bending over itself, not radiating outward.   (b) Open contour naturally shows the outward radiating order.
Fig. 4. Visualizing the effect of closed and open contours being visualized.

Instead of removing the closed contour visual information, CoC provides ways to tune visualizations, such as by rotating each circle axis to form simple visual shapes as in Fig. 5.

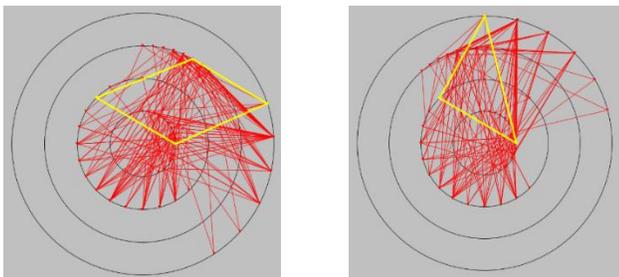

(a) After rotating clockwise of the petal width attribute of $X_4$ forming a diamond rhombus closed contour.   (b) After rotating counterclockwise, the petal length attribute of $X_3$ to form a right triangle closed contour.
Fig. 5. Iris Setosa with one-case in yellow to show open vs. closed contour.

This is done in consideration of studies in visual recognition [1] that provide both empirical and theoretical support for the perceptual advantages of closed contours over open contours. These advantages are faster detection, discrimination, neural processing, and attentional allocation. In contrast to CoC, the parallel coordinates don't have a similar closed contour option. Fig. 6 shows the benefit of axis directionality. Since each attribute can be independently oriented like how the original implementation of CoC focused on resolving crossings [19].

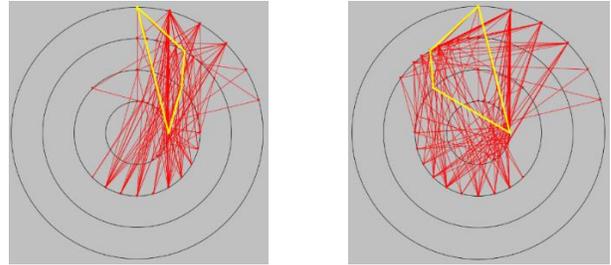

(a) reversing sepal width $X_2$ axis.   (b) reversing petal length $X_3$ axis.
Fig. 6. Impact of coordinate directionality on the CoC visualization. The highlighted case is the same as in Fig. 3.

Fig. 7 shows the same Iris data for Setosa class visualized using just one of the orders of coordinates. This shows that DTs can be used to determine an **order** of concentric circles with the attribute at the root of the DTs as the most inner circle. Moreover, coefficients of linear classifiers can be used to **spread** or **shrink** attributes on the circles to decrease crossings computationally. It can also be done interactively by software.

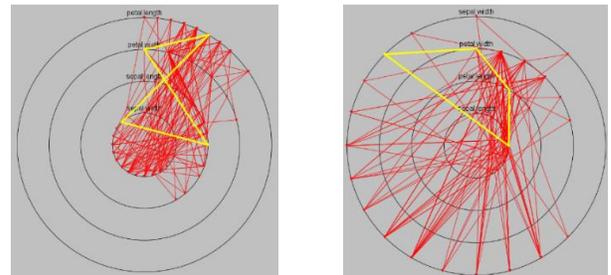

(a) DT attribute ordered inner least important with outer most important.   (b) Attribute order is by Hamiltonian path order [3]
Fig. 7. Impact of coordinate order on CoC visualization. The highlighted case is the same as in Fig. 3.

Fig. 8 shows the impact of coordinate radii being changed. Using this allows for axes to be resized relative to each other such as making one coordinate **scale** dependent on another axis. It is another way to represent data in simplified visual forms of user-desirable, closed, and perceptibly identifiable shapes.

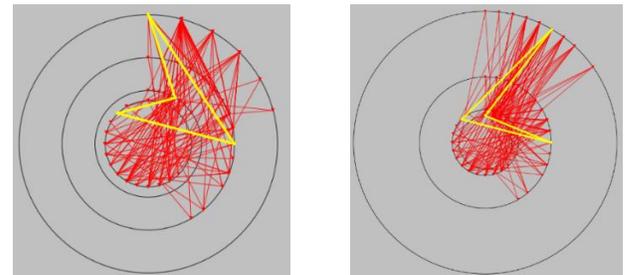

(a) Decreasing radius for the petal length forms a "triangular" shape without the case self-crossing.   (b) Decreasing the petal width radius forms another "triangular" shape without self-crossing [3]
Fig. 8. Impact of radius on CoC visualization on the case from Fig. 3.

## D. Simplifications of Concentric Coordinates

Overlap regions present problems for current classification algorithms due to being the location of many misclassifications. Figs. 9-12 illustrate the issue and resolution of case overlapping in Concentric Coordinates. Many ML classifiers including the foundational k-Nearest Neighbors (k-NN) suffer from overlap of classes in n-D visualizations. The CoC visualization allows for the visual verification of k-NN results as is described below.

**CoC with convex hull.** Fig. 9 shows the convex hulls around Iris classes in CoC. Convex hulls allow for the comparison of n-D data classes by visualization in n-D space. For instance, removing a trivial to classify region, like Setosa class simplifies visual analysis of convex hull geometry. It shows the significant known initial separation of Setosa class, particularly in the outer two attributes, petal length and petal width, in centimeter units.

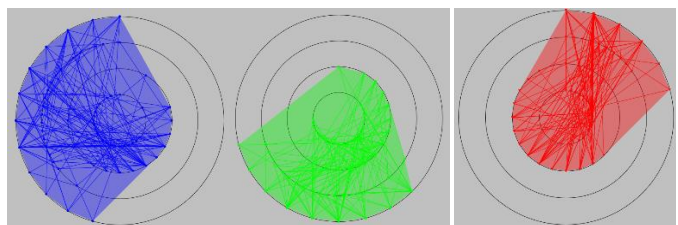

(a) Virginica.  (b) Versicolor.  (c) Setosa.
Fig. 9. Convex hulls in CoC for all three classes of the 4-D Iris data.

**CoC with straight line.** Circles can be **rotated** to make a polyline of a selected n-D point of the class a straight vertical line as shown in Fig. 10 for Setosa class Iris data. This visualization clearly shows outliers from the selected case with points furthest at the bottom and on the right marked with small black boxes. An additional benefit of a straight line case visualization is visualizing a new case as a straight line along with its **k-nearest neighbors** as presented in section II.G.

The second way to make a straight line with a given angle, and not necessarily vertical is by **adjusting radii**. Consider a 4-D point $\mathbf{x} = (x_1, x_2, x_3, x_4)$ and 4 circles with radii $R_1$ - $R_4$. We assign radius $R_1$ and compute $R_2$ - $R_4$ as follows. Compute angle $a_1 = x_1/R_1$ of $x_1$ in radians, then assign this angle to $x_2$ as $a_1 = x_2/R_2$. Next, compute $R_2 = x_2/a_1$ from it. The same way we compute $R_3 = x_3/a_1$ and $R_4 = x_4/a_1$. When $n > 4$ repeat these steps up to $n$ by computing $R_k = x_k/a_1$. The third way is keeping existing $R_1$-$R_4$ and **scale** $x_i$ differently with using some circumferences partially for values $x_i$ to accommodate scaling.

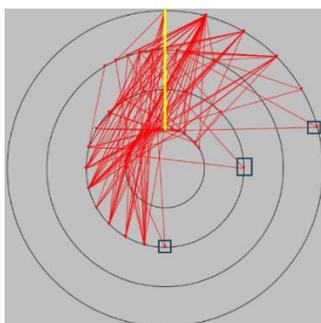

Fig. 10. Iris Setosa class in CoC with a straight line for a given case.

**CoC with synthetic mean.** For all cases selected a case is synthesized which has the mean attribute values $x_i$. Fig. 11 shows CoC with Setosa class with a synthetic mean case added. Fig. 11b shows the result of rotating the circle axes to make the mean case a straight vertical line. It made visualization of Setosa class more **compact** with more **visible patterns** including its **symmetricity** relative to the mean case, which was obscured in Fig. 11a. Additionally, Fig. 11b clearly shows outliers as lines at the bottom and sides of the CoC plot. Such simplification can be used to visually explore the separation of all three classes.

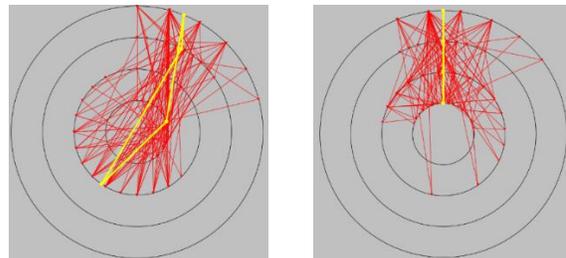

(a) Setosa class with a synthetic mean case that has a self-crossing.  (b) Effect of axes being rotated to straighten the synthetic mean case.
Fig. 11. Synthetic case as the mean of the Setosa class cases.

**CoC with frequency wideness and opacity.** The use of lines of different width and opacity to represent frequency of cases allows to make patterns more **visible** and more **fully** and **correctly** represented. Fig. 12 demonstrates this with wideness in (a) and opacity in (b). Here, the higher frequencies are presented by wider lines and less opacity, the lower frequencies are presented by thinner lines greater opacity. For instance a very visible vertical line in Fig. 12 shows the location of the most frequent cases, which is especially visible in Fig. 12b.

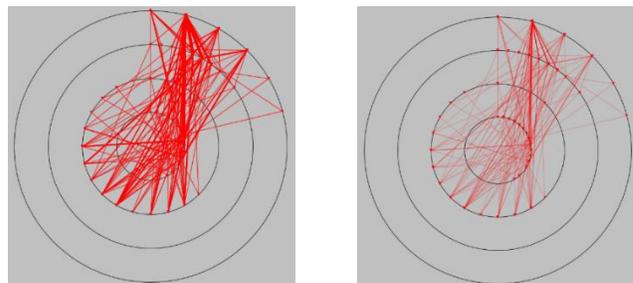

(a) Line width by frequency.  (b) Line opacity by frequency.
Fig. 12. Iris Setosa class with line width and opacity controlled by frequency.

## E. Concentric Coordinates for high dimensional data

Next, we explore capability of Concentric Coordinates for visualizing larger dimensionality of data by using pan and zoom in software to allow the concentric circles to be equidistant, while still allowing for visual exploration. Fig. 13 demonstrates the design of CoC visualization on the Iris data, 9-D Wisconsin Breast Cancer (WBC) data, and the 166-D Musk molecule data.

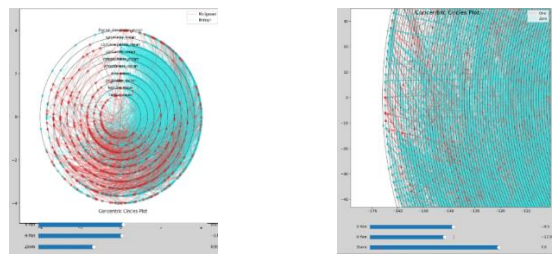

(a) WBC 9-D dataset.  (b) Musk molecule 166-D dataset.
Fig. 13. Panning and zoom controls in CoC for data of larger dimensions.

## F. Data exploration in Concentric Coordinates

First, we search for the forms of Concentric Coordinates that best reveal valuable visual patterns for data classification tasks. Followed by data exploration in those Concentric Coordinates.

**Validation of k-nearest neighbors.** One data exploration process is validation of k-nearest neighbors for classifying a given new case. Here prediction by *k*-NN for this case is marked in visualization after simplifying the case to a straight line. Visual analytics of such visualizations allows us to know when to appropriately combine classifiers for improved ensembles.

**CoC on sampled statistical distribution.** Another data exploration process with concentric coordinates is analysis of discovered visual patterns. It can be done by sampling cases that satisfy these patterns with their close statistical distribution analysis as well individual cases of interest. It also can include the analysis of the outliers of these visual patterns, mean cases of the classes and errors.

**Synthetic data**. By sampling 100 cases with each case having 10 attributes, from two normal distributions, one with a mean of 0.25 and the other with a mean of 0.75, both having a standard deviation of 1. This yields two unique data classes of 100 10-D cases each for 200 10-D cases total. We plot the first distribution in red and the second in blue. We visualize this in CoC by plotting each attribute center outwards, each attribute being mapped to the axis with 0 being the top-most point, the values increasing in clockwise orientation, axes ending at attribute max or 1. Data was normalized prior to visualization by min-max normalization resulting in a range of [0-1]. Fig. 14 shows the expected statistical separation.

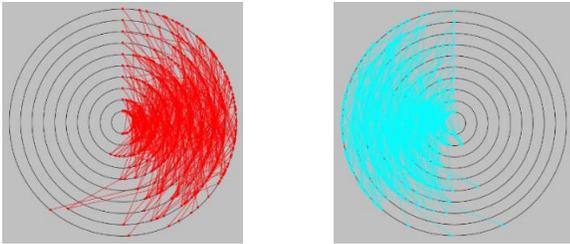

(a) Class with a lower mean.   (b) Class with a higher mean.
Fig. 14. CoC plot of sampled normal distributions with different means.

## G. Concentric Coordinates and occlusion

For data classification tasks, occlusion is a significant challenge as Fig. l5a shows Iris data. Below we present an algorithm on addressing occlusion issue for CoC. We denote this algorithm as an Occlusion Removal (OR) algorithm) algorithm with the steps below. Each n-D point in CoC is a polyline connecting points. In this algorithm we call these points nodes. The node is called pure node is only cases of single class are at that node. The node is called an overlap node if cases of 2 or classes are in that node.

**Occlusion Removal (OR) algorithm:**
**Step 1.** Find all pure nodes in each class in CoC.
**Step 2.** Count the number of cases in each pure node.
**Step 3.** Order pure nodes in decreasing order of the number of cases in the nodes.
**Step 4.** Set up a threshold for the smallest number of cases in the node.
**Step 5.** Select all nodes with the number of cases above the threshold.
**Step 6.** Mark nodes selected in step 5 with larger circles
**Step 7.** Remove all connecting lines for cases from selected nodes
**Step 8.** Build an envelope around selected points.
**Step 9.** Build a local classifier for the envelope: classify a new case to the class of the cases in the envelope if the case is in the envelope.
**Step 10.** Build a generalized DT with the local classifiers from step 9.

Where the generalized DT visualizes the relations between pure nodes (rules generated iteratively). Fig. 15 shows that result of applying steps 1-7 of this algorithm to Iris data. It is visible that that occlusion significantly decreased.

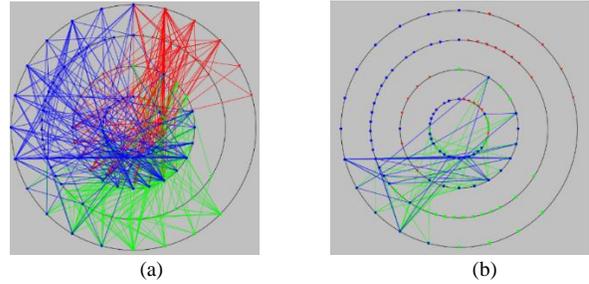

(a)                                    (b)
Fig. 15. All Iris data in CoC before (a) and after occlusion removal (b).

For the overlap area *k*-NN looks like a natural candidate due to a small number of cases. K-NN can work with a small number of cases. But the use of Euclidean distance may be the wrong one. We have the ability with CoC to fix this by defining better metrics like threshold similarity metrics [4]. How use users defined actual NN to adjust the metrics.

This visualization allows for analysis of the classification quality by *k*-NN for individual cases figure 16 shows a case from the green class which was misclassified by *k*-NN. Furthermore, results can be projected to an axis as done in [17]. Considering the green Versicolor class from Iris *k*-NN is trained on all Fisher Iris 150 cases 3 class data, other classes are visually suppressed. Fig. 16 demonstrates this visualization. CoC can be further simplified by the hiding of connecting polylines, which traverse pure regions of the plot. This allows for hiding of non-overlap regions. Fig 15 demonstrates this visualization in CoC. We can also get a difficult case from Versicolor class and visualize it with its 3-NN and 5-NN.

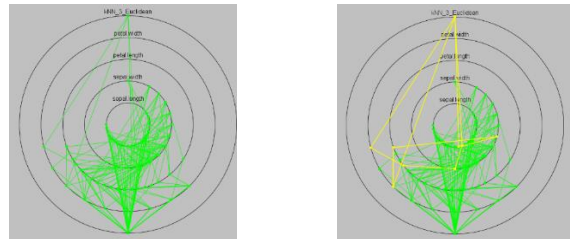

(a) *k*-NN model trained on all Iris data then classified by *k*-NN with a *k = 3* and Euclidean distance. Points going straight up were classified wrongly by *k*-NN, down are correct.   (b) Two error cases of (a) drawn in yellow to visualize the relation of errors to correct cases of the class.

Fig. 16. Singular *k*-NN classification model visualization analysis example.

## H. Concentric Coordinates and ensemble of k-NNs

The *k*-NN is one of the computational ML methods that has important advantages in which a user can understand how this algorithm predicts. Using visualized classification models like *k*-NN in an ensemble have shown to perform better than single prediction models [11]. In Section II.C we defined a way to make a case straight in CoC. Below we present how *k*-NN can be combined with CoC by visualizing a new case as a straight line as defined in section II.D.

A new case to be predicted by *k*-NN can be shown as a straight line in CoC along its *k* nearest neighbors. It can be for a single *k* like *k* = 3 or a set of *k* values, like {1,3,5,7,21}. See Fig. 17. If an ensemble of the k-NN classifiers is used. It will allow a user to visually **confirm or refute** the prediction of class label for new case(s) by *k*-NN since a user will be able to see all these nearest neighbors together without any loss of information. It also allows us to show other cases. The nearest neighbors computed by *k*-NN by common distances not necessarily are the closest ones for the user. The straight line of the case of the interest has advantage of its simplicity and a **preattentive** property so that users can concentrate on analysis of differences from of the neighboring cases relative to this straight line.

Below presents a new algorithm that is a **k-NN ensemble** denoted as **k-NNE**. It has an advantage of combining several levels of **generalization** from one-to-many neighbors. This learning process on training and validation data is as follows:

**k-NNE** algorithm:
**Step 1.** Run *k*-NN with *k* = 1 in 10-fold cross validation.
**Step 2.** Run *k*-NN in 10-fold cross validation with incremental odd *k* up to *k* = *K*, where *K* is a predefined limit.
**Step 3.** Order results of all steps (1) - (2) by average accuracy.
**Step 4.** Combine two k-NNs with the highest accuracy to vote in ensemble and compute 10-fold cross validation accuracy.
**Step 5.** Add k-NNs with next accuracies to vote until the accuracy of ensemble does not improve anymore in 10-fold cross validation.

In the case of vote ties this algorithm is adjusted by using the *k*-NNs that win overall in pair-wise comparisons of individual models as by the Copeland rule [2]. If any individual models have a high standard deviation in their accuracy as an individual model, then they should not be considered for use in ensemble.

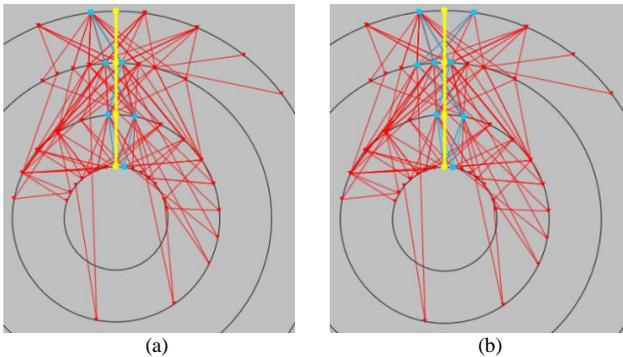

(a) (b)
Fig. 17. Synthetic case (average of Setosa class cases) in yellow with the (a) 3 and (b) 5 nearest neighbors to this case in blue, closed contours not drawn.

Fig. 18 shows larger data of 30-D Wisconsin Breast Cancer. Here we chose 15 nearest neighbors of the synthetic mean case (in yellow). This shows the benefits of straight-line and *k*-NN simultaneously. The mean case and its nearest neighbors are much simpler for visual analysis in Fig. 18b than in Fig. 18a.

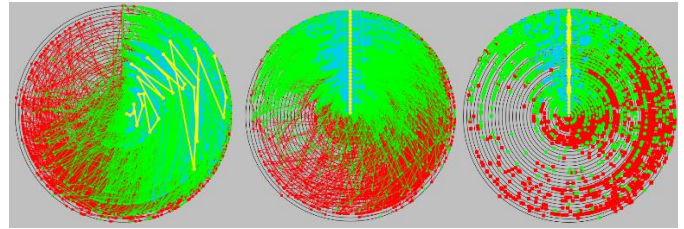

(a) CoC visualization of the synthetic mean case and its 15-NN. 
(b) CoC visualization, circles rotated so mean case is a straight-line 
(c) CoC after using the Occlusion Removal algorithm.

Fig. 18. 30-D WBC data (569 cases, 357 benign and 212 malignant) with synthetic mean benign case (yellow) and its 15 nearest neighbours in CoC.

## III. GENERALIZATION

The section below will present generalizations of Concentric Coordinates in 2-D and 3-D visualization variations. Both with significant parameterized features of radii, rotation, and ways to place circle axes in varying visual arrangements.

### A. 2-D Planar Concentric Coordinates Generalizaion

The 2-D generalization below **spreads coordinate circles** non-concentrically in the 2-D plane as shown in Fig. 19. This visualization is inspired by the decision tree visualization data flow in SPC-DT method [9]. It allows for decreasing occlusion and revealing improved visual patterns. It clearly shows distinct patterns of three Iris classes. To make patterns more visible circles can be moved by a user as it is done in SPC-DT method.

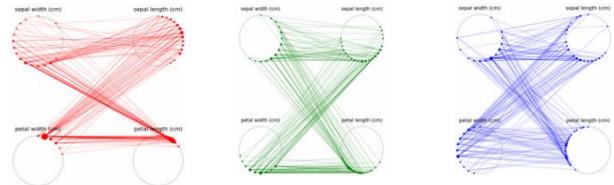

Fig. 19. CoC plot of the Iris data where the coordinates circles are spread out planarly, classes are red (Setosa), green (Versicolor), and blue (Virginica).

### B. 3-D Spatial Concentric Coordinates Generalization

The following section presents several generalizations of the Concentric Coordinates visualized in 3-D.

**3-D Concentric Coordinates.** In Fig. 20, the generalization is produced by substituting concentric circles with concentric spheres. In 3-D generalizations visualization can use any closed contours that are a convex 3-D shape topologically equivalent to the sphere, like an ellipsoid. Axes in 3-D can be resized, rearranged, reordered, and rotated, just as was shown in 2-D. In general, 3-D visualization has an advantage of selecting a view where the classes are most visibly well-separated.

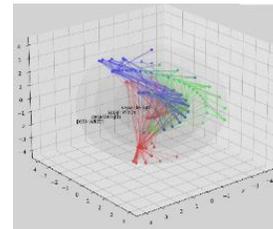

Fig. 20. 3-D Concentric Coordinates of Iris data using spherical axes.

**3-D Stacked Coordinates.** In 3-D Stacked Coordinates circles are put to 3-D non-concentrically. Fig. 20 shows circles stacked cylindrically. Fig. 21. circles stacked expanding and shrinking.

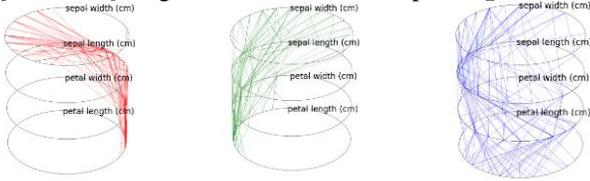

(a) Three classes of Iris data in CoC with spreading coordinates on Z-axis.

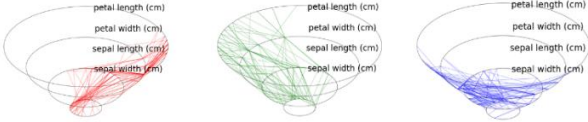

(b) Three classes of Iris data in expanding coordinates on Z-axis.

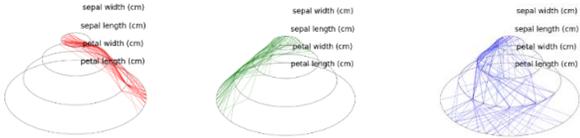

(c) Three classes of Iris data in shrinking coordinates on Z-axis.
Fig. 21. 3-D stacked axes with decreasing circle axis radii for each attribute. (red-setosa, green- versicolor, blue- virginica).

### C. Generalized Iterative Classifier

Several machine learning algorithms have an **iterative** nature. These algorithms classify subsets of data that are associated with a **network of nodes**, Decision tree and Neural Networks are examples of such iterative classifiers. Some of these algorithms have advantages of fast computation and interpretability. Below we define a Generalized Iterative Classifier (GIC) and explore it. The **Generalized Iterative Classifier** (**GIC**) is a triplet formally defined as:

$$G = <\{A_{classifier}\}, I_{max}, \{\rho_{threshold}\}>$$

where $\{A_{classifier}\}$ is a set of classifier algorithms, $I_{max}$ is the limit on the number of iterations, and $\rho_{threshold}$ is a set of purity of classification thresholds. Each node is associated with an algorithm $A_{classifier}$ and a threefold $\rho_{threshold}$, that can be the same or different for different nodes.

GIC allows for constructing of a regular and a Generalized Decision Tree (GDT) [17] as a predictive ML classification model and as well as other iterative classifiers. Examples: <Logistic Regression, 3, 0.95>, <Support Vector Machine (Linear), 5, 1.00>, or <Multilinear Perceptron, 10, 0.95>. Note that these examples all show using the same classifier per node, this is not strictly required. The classifier used can be changed at each node i.e. using a more complex SVM kernel from linear to polynomial to high powers thereof as the iterations converge.

Another example is of a **single-attribute classifier** (**SAC**) algorithm. This algorithm builds iteratively a set of rules $R(\mathbf{x})$ in one of the following forms (1) - (3).

$$R(\mathbf{x}) = 1 \Leftrightarrow T_1 \leq F(x_i) \leq T_2 \text{ \& } \mathbf{x} \in \text{Class L} \quad (1)$$

$$R(\mathbf{x}) = 1 \Leftrightarrow T_1 \leq F(x_i) \text{ \& } \mathbf{x} \in \text{Class L} \quad (2)$$

$$R(\mathbf{x}) = 1 \Leftrightarrow F(x_i) \geq T_2 \text{ \& } \mathbf{x} \in \text{Class L} \quad (3)$$

where $\mathbf{x}$ is an n-D point $\mathbf{x} = (x_1, x_2, \ldots, x_n)$, $F$ returns the attribute value $x_i$ for n-D point $\mathbf{x}$, and $T_1$ and $T_2$ are thresholds. For instance, for L = 1, $i = 3$, $T_1 = 0.3$ and $T_2 = 0.5$, $R(\mathbf{x}) = 1$ means that every n-D point $\mathbf{x}$ from a given set of n-D points S belongs to class 1 when $0.3 \leq F(x_3) \leq 0.5$. We will call these cases **pure cases** of class L = 1. For some n-D points $\mathbf{x}$ and $\mathbf{y}$ from different classes we may have a situation when $F(x_i) = F(y_i)$. In this case we call $\mathbf{x}$ and $\mathbf{y}$ **overlap cases**.

**Iteration steps.** At Step 1 the algorithm finds all overlap cases OL1. At Step 2 the algorithm works only with cases OL1 and searches for overlap cases in OL1 to produce the overlap cases OL2. At the next steps overlap cases OL3, OL4, etc. are produced. This process is finite because we either do not find any pure cases on the next step or find them and will get a smaller set of overlap cases to proceed with. With the total finite number of cases in the dataset this process is finite too.

**Statement 1 (Convergence of Single-attribute Classifier).** For every finite set $S$ of labeled n-D points a single-attribute classifier will **stop** classifying after a finite iteration of steps.

Below is an example of how a set of overlap cases shrinks with iteration. Consider an example of a dataset with 1000 n-D points of two classes with 500 cases each. The single-attribute classifier algorithm searches for rules (1) - (3). It is possible that no single n-D point with such property exists. It means that the algorithm finishes in one step without classifying any n-D point. If some n-D points, say 200 n-D points, are classified at step 1, then the algorithm attempts to classify the remaining 800 n-D points. Removal of 200 pure cases can create new pure cases in the 800 cases as the example below shows.

Let at the first iteration we have two n-D pints $\mathbf{a}$ from class 1 and $\mathbf{b}$ from class 2 and both with $x_1 = 2$. We also have $x_2 = 4$ for $\mathbf{a}$ and no n-D points of class 2 with $x_2 = 4$. This led to the exclusion of n-D point $\mathbf{a}$ as a pure case at step 1. Thus, at step 2 we have only $\mathbf{b}$ with $x_1 = 2$ and now we can exclude it because its overlap with $\mathbf{a}$ is not present anymore. It is the same process as used in building decision trees and generalized decision trees in [16] iteratively from layer to layer. If such $\mathbf{a}$ and $\mathbf{b}$ do not exist, the algorithm stops. Otherwise, it continues to further steps. Having a finite set of n-D points this process will stop after a finite number of steps when all n-D points are tested.

This algorithm can end up classifying only a subset of data because some cases can still have overlap cases. The fact that the process is finite allows us to analyze the cases that were not classified being the overlap cases at each step. The overlap cases of the final step require a separate classification method. These cases are also strong candidates for the worst-case validation and test data for a classifier [14]. Experiments show that for Wisconsin Breast Cancer data the number of those overlap cases is less than 10% of the original data, so they can be analyzed visually on CoC and other GLC with much less occlusion than for the whole dataset.

Below we consider an iterative linear classifier, which is in contrast with a single-attribute classifier, can classify all cases in the same iterating process of classifying at the next step only overlap cases when the dataset has no equal cases in different classes. The reason is that for a linear classifier rule

$$R(\mathbf{x}) = 1 \Leftrightarrow T_1 \leq F(\mathbf{x}_i \,\&\, \mathbf{x} \in \text{Class L}) \quad (4)$$

has $F(\mathbf{x})$ not $F(x_i)$ as in (2) for a single-attribute classifier, where $F(\mathbf{x}) = a_1x_1 + a_2x_2 + \ldots + a_nx_n$ is a linear function. Note that this does not avoid overfitting in being able to classify any case.

**Statement 2 (Convergence of Iterative linear Classifier).** For every finite set S of labeled n-D points that does not contain any duplicate n-D points $\mathbf{x} = \mathbf{y}$ of different classes, an iterative linear classifier G exists that **classifies** all n-D cases of S after some finite integer count of iterated steps or reaches the $I_{max}$ maximum number of iterations.

The proof follows from (4). Consider two overlap cases $\mathbf{a}$ and $\mathbf{b}$ from two classes with $F(\mathbf{a}_i) < F(\mathbf{b})$ then we can find $T_1$ such that $F(\mathbf{b}) < T_1 < F(\mathbf{b})$, separate $\mathbf{a}$ and $\mathbf{b}$ and classify them correctly by using this threshold in (4). With a finite number of cases in the dataset the number of such overlap pairs of cases to be separated is finite too. To avoid overfitting the developed JTabViz software defaults to a 5% minimum pure-region region size threshold. This result is similar in nature to the classical convergence proofs of Perceptron Learning for separable cases but extended to iterative decision function refinement [13].

## IV. CONCLUSIONS AND FUTURE WORK

Predictions made by AI/ML systems for high-risk domains must be precise, visualizable, and well-understood by subject experts who must deploy such systems. This is required to guarantee safety and reliability. The expanded GLC approach of CoC was inspired by recent successes with related GLCs. Utilization of the CoC along with alternative GLC visualizations can greatly benefit visual pattern discovery, therefore the HITL must appropriately choose what method to proceed with for each AI/ML data and selected subsets to balance model accuracy at the case and global levels with generalizability to unseen data. Since feature engineering approaches can be white-box or grey-box, being only explained after model design, or quasi-box [7] being only explained in results visually. Instead, our work integrates human visualization with algorithmic computation. This allows for visual discovery of generalized data properties to classify the data with the Divide and Classify process.

The advantage of the Concentric Coordinates approach is in allowing adding more circles to represent additional attributes within computational and visual limitation. CoC allows for a compact usage of visual space by controlling the space between circles. Moreover, the circle axes represent n-D data on expanding sized circles with this being controlled by the size of the circles. Additional parameterized features are available to tune the perceptibility of visualized n-D data for engineering of simple visual patterns. Future work should focus on further exploration of the geometric and topological properties of ML/AI datasets and seek isolation of principal attributes to visualize with which to classify larger datasets such as MNIST. Approaches will require continued work with interactive data transformation directly in n-D spaces to optimize problematic cases to classification directly. Using CoC further to visualize decision flow pathways in Neural Networks may allow for visual analysis for the task of neural network simplification. Lastly, combining this work's feature engineering with visually driven synthetic data generation tools already developed [18] to further tune models in the creation process, by data geometric transformations directly in visualized n-D space.